\documentclass[longnamesfirst,apjl]{emulateapj}
\usepackage{apjfonts,graphicx}
\def\apgt{\ {\raise-.5ex\hbox{$\buildrel>\over\sim$}}\ }
\def\aplt{\ {\raise-.5ex\hbox{$\buildrel<\over\sim$}}\ }

\tighten

\shorttitle{CORE FORMATION}
\shortauthors{MERRITT ET AL.}

\newcommand{\lap}{\lesssim}
\newcommand{\gap}{\gtrsim}

\newcommand{\msun}{M_\odot}

\newcommand{\beq}{\begin{equation}}
\newcommand{\eeq}{\end{equation}}
\newcommand{\mh}{m_{BH}}
\newcommand{\Mh}{M_{BH}}

\newenvironment{inlinefigure}{
\def\@captype{figure}
\noindent\begin{minipage}{0.95\linewidth}\begin{center}}
{\end{center}\end{minipage}\smallskip}

\begin{document}

\title{Core Formation by a Population of Massive Remnants}

\author{David Merritt           \altaffilmark{1},
        Slawomir Piatek         \altaffilmark{2},
        Simon Portegies Zwart   \altaffilmark{3}, and
        Marc Hemsendorf         \altaffilmark{4}}

\altaffiltext{1}{Department of Physics, Rochester Institute of Technology,
Rochester, NY 14623}

\altaffiltext{2}{Department of Physics, New Jersey Institute of Technology,
Newark, NJ 07102}

\altaffiltext{3}{Astronomical Institute ``Anton Pannekoek''
and Institute for Computer Science, University of Amsterdam, 
Kruislaan 403, Amsterdam, The Netherlands}

\altaffiltext{4}{Department of Physics and Astronomy, Rutgers University,
New Brunswick, NJ 08903}

\begin{abstract}
Core radii of globular clusters in the Large and Small Magellanic
Clouds show an increasing trend with age.  We propose that this trend
is a dynamical effect resulting from the accumulation of massive
stars and stellar-mass black holes at the cluster centers.  The black
holes are remnants of stars with initial masses exceeding $\sim
20-25\msun$; as their orbits decay by dynamical friction, they heat
the stellar background and create a core.  
Using analytical estimates and $N$-body experiments, we show that
the sizes of the cores so produced and their growth rates are consistent 
with what is observed.  
We propose that this mechanism is responsible for the formation of
cores in all globular clusters and possibly in other systems as well.

\end{abstract}


\section{Introduction}

An enduring problem is the origin of {\it cores},
regions near the center of a stellar or dark matter system 
where the density is nearly constant.  
Resolved cores clearly exist in some stellar systems, 
e. g. globular clusters \citep{Harris:96}.
In other systems, such as early-type galaxies, 
cores were long believed to be generic
but were later shown to be artifacts of the seeing
\citep{Schweizer:79}.  
Nevertheless a few elliptical galaxies do exhibit bona-fide 
cores \citep{Lauer:02} while many others show a central
density that rises only very slowly toward the center \citep{MF:95}.
Density profiles of structures that form from gravitational clustering
of density perturbations in an expanding universe are believed to 
lack cores \citep{Power:03}, although there is evidence for dark
matter cores in the rotation curves of some late-type galaxies
(e.g. \cite{Jimenez:03}).

The existence of a core is usually deemed to
require a special explanation.  For instance, galaxy cores may form when
binary black holes eject stars via the gravitational slingshot
\citep{Ebisuzaki:91}.

A useful sample for testing theories of core formation is the ensemble of globular
clusters (GCs) around the Large and Small Magellanic Clouds (LMC/SMC).
These clusters have masses similar to those of Galactic GCs, but many
are much younger, with ages that range from $10^6$ -- $10^{10}$ yr.
Furthermore ground-based \citep{Elson:89,Elson:91,Elson:92} and HST
\citep{Mackey:03a,Mackey:03b} observations reveal a clear trend of core radius
with age: while young clusters ($\tau\ll 10^8$ yr) have
core radii consistent with zero, clusters older than $\sim 10^9$ yr
exhibit the full range of core sizes seen in Galactic GCs, $0$ pc
$\lap r_c\lap 10$ pc (Figure~\ref{fig:mackey}).  The maximum core
radius observed in the LMC/SMC GCs is an increasing function of age
and is given roughly by $r_{c}\approx 2.25{\rm pc}\log_{10}\tau_{\rm yr} -
14.5$.
Attempts to explain the core radius
evolution in terms of stellar mass loss \citep{Elson:91}, a primordial
population of binary stars, or time-varying tidal fields
\citep{Wilkinson:03} have met with limited success.  The difficulty is
to find a mechanism that can produce substantial changes in the
central structure of a GC on time scales as short as a few hundred Myr,
while leaving the large-scale structure of the cluster intact.

In this paper, we describe a new mechanism for the formation of 
GC cores and their evolution with time.  
Massive stars and their black hole remnants
sink to the center of a GC due to dynamical friction against the
less massive stars.
The energy transferred to the stars during this process, and
during the three- and higher-$N$ encounters between the black holes
that follow,
has the effect of displacing the stars and creating a core.
The rate of core growth implied
by this model is consistent with the observed dependence of core size
on age in the LMC/SMC clusters.

\begin{inlinefigure}
\begin{center}
\resizebox{0.85\textwidth}{!}{\includegraphics{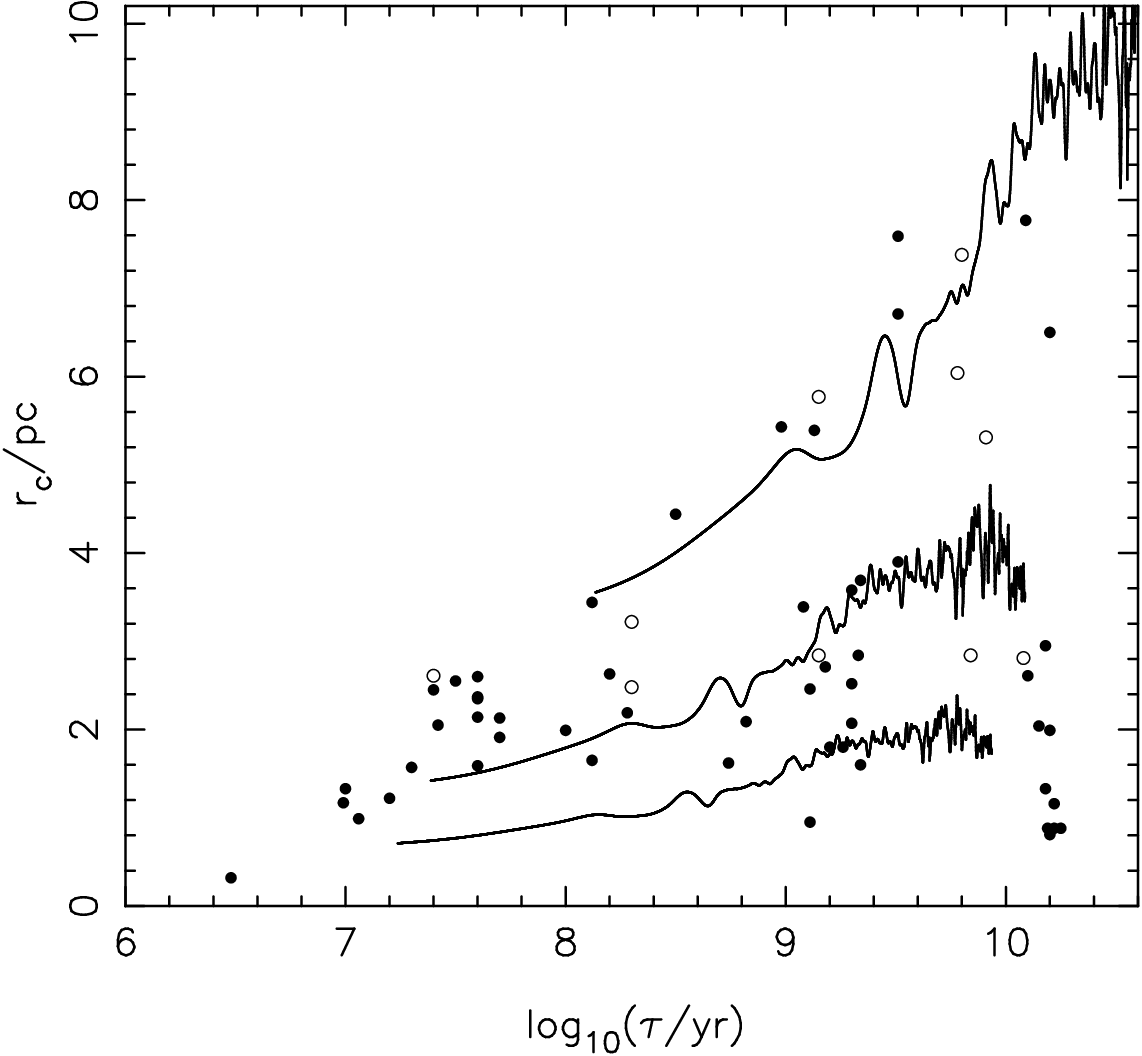}}
\end{center}
\figcaption{\label{fig:mackey} Core radius  versus age for LMC 
and SMC GCs
from the samples of Mackey \& Gilmore (2003a,b).  
Lines show core radius evolution from the $N$-body simulations 
with initial cusp slope $\gamma=1$ and three different scalings
to physical units; see text for details.  }
\end{inlinefigure}

\section{Core Formation Timescales}

Consider a gravitationally bound stellar system in which 
most of the mass is in the form of stars of mass $m$, 
but which also contains a subpopulation of more massive objects
with masses $\mh$.
The orbits of the more massive objects decay 
due to dynamical friction.
Assume that the stellar density profile is initially
a power-law in radius, $\rho(r)=K(r/a)^{-\gamma}$,
$K=(3-\gamma)M/4\pi a^3$ with $M$ the total stellar
mass and $a$ the density scale length; the expression
for $K$ assumes that the density follows a Dehnen (1993)
law outside of the stellar cusp, i.e. $\rho\sim r^{-4}$
at large $r$.
The effective radius (the radius containing $1/2$ of the
mass in projection) is related to $a$ via
$R_e/a\approx (1.8,1.3,1.0)$ for $\gamma=(1,1.5,2)$. 

Due to the high central concentration of the mass,
the orbits of the massive particles will rapidly circularize as they
receive nearly-impulsive velocity changes near pericenter.  
Once circular, orbits shrink at a
rate that can be computed by equating the torque from dynamical
friction with the rate of change of orbital angular momentum.  We
adopt the usual approximation \citep{Spitzer:87} in which the
frictional force is produced by stars with velocities less than the
orbital velocity of the massive object.  The rate of change of the
orbital radius, assuming a fixed and isotropic stellar background, is
then
\begin{eqnarray}
{dr\over dt} &=& -2{(3-\gamma)\over 4-\gamma}\sqrt{GM\over a}{\mh\over M} 
\ln\Lambda \left({r\over a}\right)^{\gamma/2-2}F(\gamma), \\
\label{eq:drdt}
F(\gamma) &=& {2^\beta\over \sqrt{2\pi}} 
{\Gamma(\beta)\over\Gamma(\beta-3/2)}(2-\gamma)^{-\gamma/(2-\gamma)}\times
\nonumber \\
& & \int_0^1 dy\ y^{1/2} \left(y+{2\over 2-\gamma}\right)^{-\beta},\nonumber
\end{eqnarray}
where $\beta=(6-\gamma)/2(2-\gamma)$ and $\ln\Lambda$ is the Coulomb
logarithm, roughly equal to 6.6 \citep{Spinnato:03}.  
For $\gamma=(1.0,1.5,2.0)$, $F=(0.193,0.302,0.427)$.  Equation
(1) implies that the massive object comes to rest at the center of the
stellar system in a time 
\beq \Delta t \approx 0.2 \sqrt{a^3\over GM}
{M\over \mh} \left({r_i\over a}\right)^{(6-\gamma)/2}
\label{eq:deltat}
\eeq with $r_i$ the initial orbital radius; the leading
coefficient depends weakly on $\gamma$.  Equation (\ref{eq:deltat})
can be written 
\beq 
\Delta t \approx 3\times 10^9 {\rm yr}\
a_{10}^{3/2} M_5^{1/2} m_{BH,10}^{-1} \left({r_i\over
a}\right)^{(6-\gamma)/2} 
\eeq 
with $a_{10}$ the density scale length in units of $10$ pc
(e.g. Figure 1 of \cite{Bergh:91}), 
$M_5=M/10^5\msun$, and $m_{BH,10}=\mh/10\msun$,
the approximate masses of black hole remnants of stars with initial
masses exceeding $\sim 20-25\msun$ \citep{Maeder:92,Zwart:97}.  This
time is of the same order as the time ($\sim 10^9$ yr) over which core
expansion is observed to take place (\cite{Mackey:03a,Mackey:03b},
Fig. 1).

To estimate the effect of the massive remnants on the stellar density profile, 
consider the evolution of an
ensemble of massive particles in a stellar system with
initial density profile $\rho\sim r^{-2}$.
The energy released as one particle spirals in from radius
$r_i$ to $r_f$ is $2\mh\sigma^2\ln(r_i/r_f)$,
with $\sigma$ the 1D stellar velocity dispersion.
Decay will halt when the massive particles form a self-gravitating
system of radius $\sim G\Mh/\sigma^2$ with $\Mh=\sum\mh$.
Equating the energy released during in-fall with the energy
of the stellar matter initially within $r_c$, the
``core radius,'' gives
\beq
r_c\approx {2G\Mh\over\sigma^2}\ln\left({r_i\sigma^2\over G\Mh}\right).
\eeq
Most of the massive particles that deposit their energy
within $r_c$ will come from radii $r_i\approx {\rm a\ few}\times r_c$,
implying $r_c\approx {\rm several}\times G\Mh/\sigma^2$
and a displaced stellar mass of $\sim {\rm several}\times \Mh$.
If $\Mh\approx 10^{-2}M$ \citep{Zwart:00},
then $r_c/a\approx {\rm several}\times 2 \Mh/M$ and
the core radius is roughly $10\%$ of the effective radius.

Evolution continues as the massive particles form binaries and
begin to engage in three-body interactions with other massive
particles.  
These superelastic encounters will eventually eject
most or all of the massive particles from the cluster.
Assume that this
ejection occurs via the cumulative effect of many encounters, 
such that almost all of the binding energy so
released can find its way into the stellar system as the particle
spirals back into the core.  The energy released by a single binary in
shrinking to a separation such that its orbital velocity equals the
escape velocity from the core is $\sim\mh\sigma^2\ln(4\Mh/M)$.  If all
of the massive particles find themselves in such binaries before their
final ejection and if most of their energy is deposited near the
center of the stellar system, the additional core mass will be
\beq 
     M_c\approx \Mh\ln\left({M\over\Mh}\right) 
\label{eq:addit}
\eeq 
e.g. $\sim 5\Mh$ for $M/\Mh=100$, similar to the
mass displaced by the initial infall.  
The additional mass
displacement takes place over a much longer time scale however and
additional processes (e.g. core collapse) may compete with it.

\begin{inlinefigure}
\begin{center}
\resizebox{\textwidth}{!}{\includegraphics{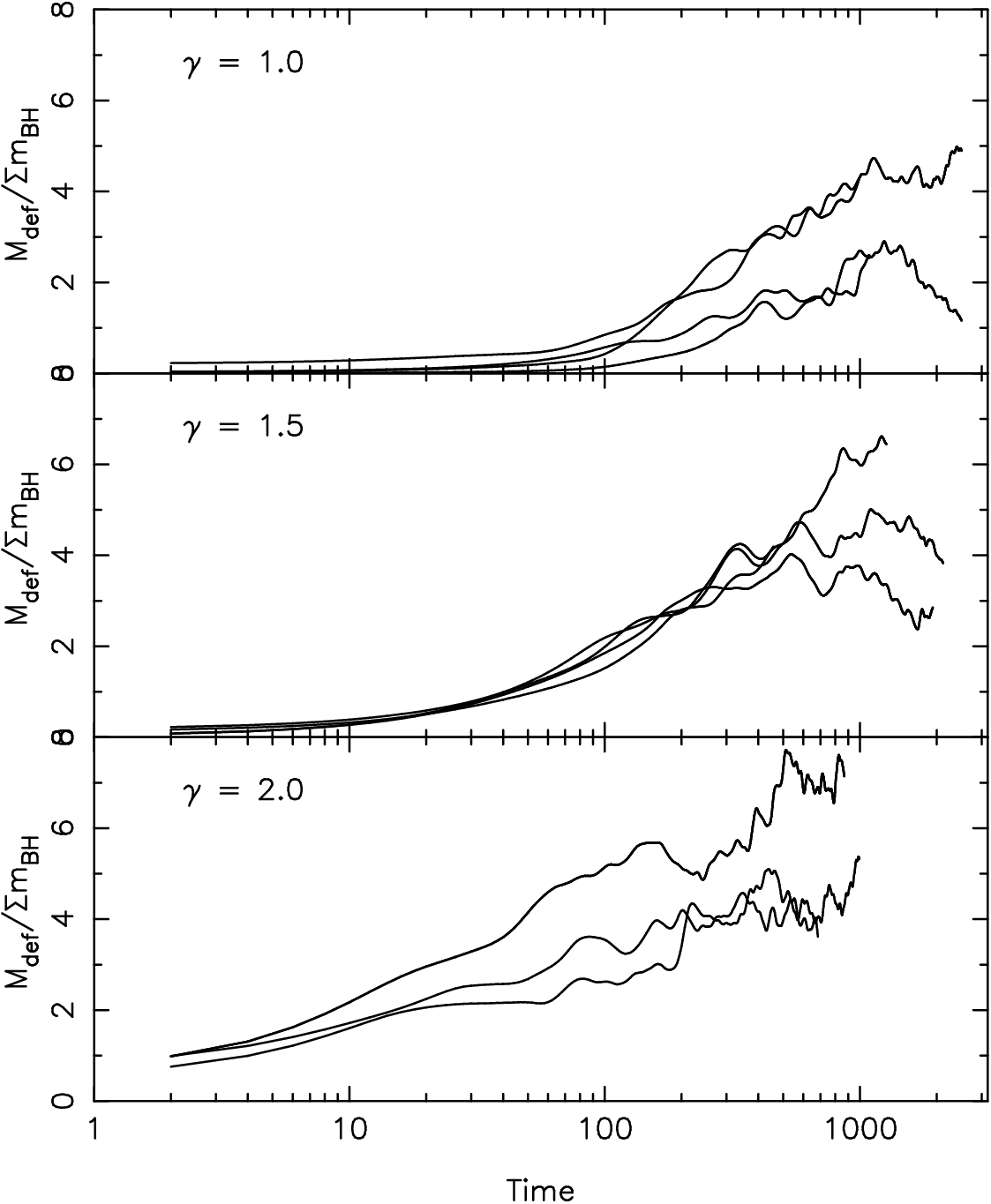}}
\end{center}
\figcaption{\label{fig:mdef}Evolution of the mass deficit,
from the $N$-body experiments with $N_{BH}=10$.  
Data were smoothed with cubic splines.}
\end{inlinefigure}

\section{$N$-body Simulations}

We used $N$-body simulations to test the core formation mechanism
described above.  Initial conditions were designed to represent GCs in
which $1\%$ of the total mass is initially in the form of massive objects,
either stars or their black hole remnants.
Integrations were carried out using NBODY6++, a high-precision,
parallel, fourth-order direct force integrator which implements
coordinate regularization for close encounters \citep{Spurzem:99}.
Particles had one
of two masses, representing either black holes ($\mh$) or stars ($m$).
The number of particles representing black holes was
$N_{BH}=(4,10,20)$
and the ratio of $\mh$ to $m$ ranged from 10 to 25.  
Most of the $N$-body experiments used $N=10^4$ particles.  
We concentrate here on the results obtained with $N_{BH}=10$
and $\mh/m=10$; results obtained with other values of
$N_{BH}$ were consistent.
All particles were initially distributed according to
Dehnen's (1993) density law, $\rho(r)=((3-\gamma)M/4\pi
a^3)/\xi^{\gamma}/(1+\xi)^{4-\gamma}, \xi=r/a$.  
The logarithmic slope of the central density cusp is specified
via the parameter $\gamma$.
Initial velocities were generated
assuming isotropy; positions and velocities of the massive
particles were distributed in the same way as the stars, so that the
initial conditions represented a cluster in which the massive objects
had not yet begun to segregate spatially with respect to the 
lighter stars.

The $1\%$ mass fraction in massive particles was based on
a Scalo (1986) mass distribution with lower and upper mass limits 
of $0.1\,\msun$ and $100\,\msun$ respectively.
With such a mass function, about 0.071\% of the stars are more 
massive than $20\,\msun$ and $0.045$\% are more massive than $25\,\msun$.
A star cluster containing $N_\star$ stars thus produces 
$\sim6\times 10^{-4} N_\star$ black holes.  
Known Galactic black holes have masses $\mh$ between
$6\,\msun$ and $18\,\msun$ \citep{Timmes:96}.
Adopting an average black hole mass of $10\,\msun$ then 
results in a total black hole mass of
$\sim 6 \times 10^{-3}M$.

The decision to use just two mass groups -- 
clearly an idealization of the true situation -- 
was made for two reasons.
First, the interpretation of the $N$-body results is greatly
simplified in such a model.
Second, it is not clear what a better choice for the
initial spectrum of masses would be.
The distribution of black hole masses produced by stars with
$m\gap 25\msun$ is uncertain \citep{Timmes:96};
some Galactic black holes may have masses as
low as $\sim 3\msun$ \citep{White:96}, 
close to the maximum probable masses of neutron stars.
However this may be a selection effect: low-mass binary
systems tend to be selected due to their longer lifetimes.
The remnant mass range between $\sim 1\msun$ and $\sim 3\msun$
is occupied by neutron stars, 
but there is evidence that neutron stars
receive larger kicks at birth than black holes
\citep{Lyne:94} and may be ejected.
(Portegies Zwart \& McMillan (2000) estimate that
$\sim 10\%$ of black holes are ejected from GCs 
by formation kicks.)
In summary, the initial spectrum of masses in a GC shortly after
its formation is poorly known.
Future studies will attempt to include a realistic 
treatment of stellar physics, primordial binaries,
star formation, and other processes that affect the 
initial mass spectrum and the initial spatial distribution
of different mass groups.

If the core radius is defined as the radius at which the 
projected density falls to $1/2$ of
its central value, Dehnen models with $\gamma\ge 1$ have
$r_c=0$.
Any core that appears in these models must therefore be a
result of dynamical evolution.  

Henceforth we adopt units in which $G=a=M=1$.
The corresponding unit of time is 
\begin{equation}
\left[T\right] = \left[{GM\over a^3}\right]^{-1/2} = 
1.44\times 10^6\ {\rm yr}\ M_{5}^{-1/2}a_{10}^{3/2}
\label{eq:tscale}
\end{equation}
where $M_5$ is the cluster mass in units of $10^5\msun$ and
$a_{10}$ is the cluster scale length in units of 10 pc.
The effective radius $R_e$, defined as the radius containing
$1/2$ of the light particles seen in projection, is 
$R_e\approx (1.8,1.3,1.0)$ in model units ($a=1$) 
for $\gamma=(1,1.5,2)$.
The time scaling of equation (\ref{eq:tscale})
is not correct for processes
whose rates depend on the masses of individual stars or
black holes, since our models have fewer stars than
real GCs.
The most important of these processes for our purposes
are black hole-star interactions, which are responsible
for the orbital decay of the black holes and the growth
of the core.
This decay occurs in our simulations at a rate that is
$\sim N_\bullet/N_{BH}$ times faster than implied by the 
scaling of equation (\ref{eq:tscale}), with $N_\bullet$ the
true number of black holes in a GC.
Assuming a Scalo initial mass function as above,
this factor may be written $\sim 6.0(N_\star/10^5)/(N_{BH}/10)$
with $N_\star$ the true number of stars in a GC.

As discussed above, we expect the stellar mass displaced 
by the massive particles to scale roughly with 
$\Mh$.
One way to illustrate this is via the {\it mass deficit},
defined as in \cite{Milos:02}:
it is the mass difference between the initial stellar density
$\rho(r,0)$ and the density at time $t$, 
integrated from the origin out to the
radius at which $\rho(r,t)$ first exceeds $\rho(r,0)$.
The mass deficit is a measure of the core mass.
Figure 2 shows $M_{\rm def}(t)/\Mh$ for the $N$-body experiments.
The density center was computed via the Casertano-Hut (1985) algorithm.
The black holes displace a mass in stars of order 
$2-8$ times their own mass; the larger values correspond to
the larger values of $\gamma$ although there is considerable
scatter from experiment to experiment for a given $\gamma$.
The results for $\gamma=2$ are consistent with the analytic arguments
presented above, which implied a core mass of a few
times $\Mh$ after a time in model units of $\sim 0.2 M/\mh\approx 20$
(cf. equation~\ref{eq:deltat}) followed by a slower displacement
of a similar mass as the black hole particles engage in
three-body interactions (cf. equation~\ref{eq:addit}).

Some of the $N$-body simulations show a decrease in the core
radius after $t \approx 10^3$
(Figure 2, 3). 
By this time, the majority of the black holes have been ejected.  
Our simulated clusters are then
effectively reduced to equal-mass systems, which take about 15 half
mass relaxation times to experience core collapse \citep{Spitzer:87}.
The two-body relaxation time is roughly $T_R \approx 0.2 T_D N/\ln N \approx
200$ in our $N$-body models, with $T_D$ the crossing time.
It is therefore not surprising that, once the black
holes are ejected, the cluster core shrinks again on a time scale of
$\sim 10^3$ time units.

About one-tenth of the known globular clusters in the Galaxy have
vanishingly small cores and are inferred to be in a state of core 
collapse \citep{Harris:96}.
We note here the large number ($\sim 10$) of GCs in
Figure\,\ref{fig:mackey} with small or zero core radii;
this may indicate that a much larger fraction ($\sim 80\%$) of 
the LMC globular clusters are on their way to core collapse. 
We predict that these clusters have lost most or all of their 
black holes, while the $\sim$ two old clusters in Figure\,\ref{fig:mackey}
with substantial cores still contain a few stellar mass black holes in
their cores.
We find no indication from their structural parameters that these
two clusters differ systematically from the other clusters
in Figure 1.

Figure\,\ref{fig:rc} shows the evolution of the core radii 
in these simulations.
Computation of $r_c$ was based on its standard
definition as the projected radius at which the surface
density falls to $1/2$ of its central value.
Projected densities were computed via a kernel estimator
\citep{Merritt:94,Merritt:04}.
To reduce the noise, values of $r_c$ from all experiments with the same
$\gamma$ and with $N_{BH}=10$ were averaged together.
Figure 3 shows that core sizes increase roughly as the logarithm
of the time, consistent with the time dependence of
the upper envelope of Figure 1,
and reach values at the end of the simulations of
$\sim 10\%$ of the half-mass radius.

Based on equation (6) and the discussion following,
the conversion factor from model time units to physical
time units is approximately
\beq
8.9\times 10^6\ {\rm yr}\ M_5^{-1/2}a_{10}^{3/2}N_{\star,5}
\label{eq:tscale2}
\eeq
with $N_{\star,5}$ the number of stars in the GC in units of
$10^5$.
This scaling was used to plot three curves in Figure 1:
with $M_5=N_{\star,5}=2$, $a_{10}=0.5$ (bottom),
$M_5=N_{\star,5}=0.5$, $a_{10}=1$ (middle), 
and $M_5=N_{\star,5}=1$, $a_{10}=2.5$ (top).
The curves in Figure 1 were taken from the experiments with 
$\gamma=1$;
the experiments with $\gamma=1.5$ and $2$ give similar results
(note that $R_e/a$ varies by a factor $\sim 2$
from $\gamma=1$ to $\gamma=2$, hence $r_c/R_e$ varies
less than $r_c/a$ in Figure 3).
The logarithmic time dependence of the upper envelope of the $r_c$
distribution is well reproduced, and with appropriate
(and reasonable) scaling, points below the envelope
can also be matched.
As noted above, the smaller core radii that begin to appear in
SMC/LMC clusters with $\tau\gap 10^9$ yr are plausibly
due to evolution toward core collapse in these clusters,
as seen also in some of the simulations.

\begin{inlinefigure}
\begin{center}
\resizebox{\textwidth}{!}{\includegraphics{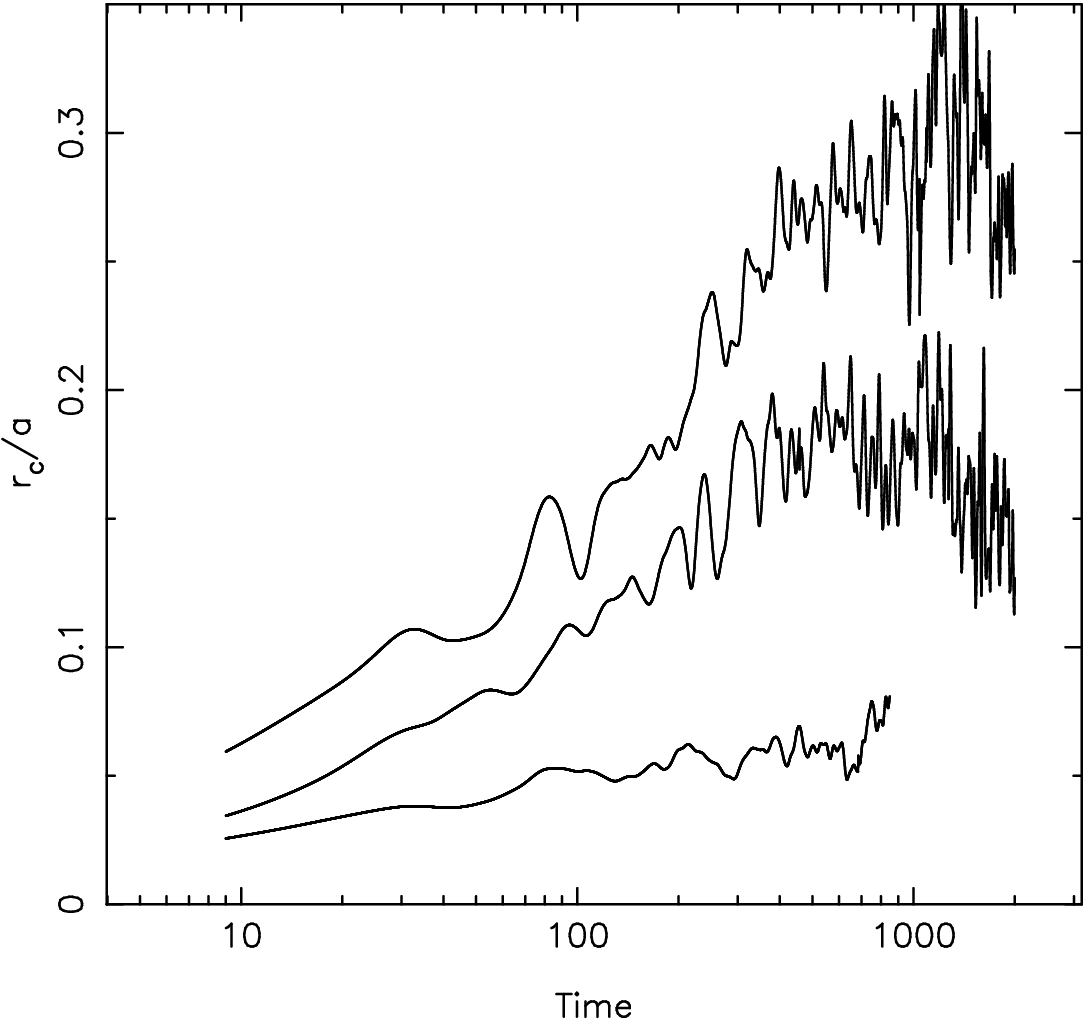}}
\end{center}
\figcaption{\label{fig:rc}Evolution of the core radius,
defined as the radius at which the projected density falls to
one-half of its central value.
Each curve is the average of the various experiments at the specified
value of $\gamma$, with $\gamma=0.5$ (top), $\gamma=1$ (middle)
and $\gamma=2$ (bottom).
Vertical axis is in units such that the Dehnen-model scale
length $a=1$; see text for conversion factors from $a$ to $R_e$.
}
\end{inlinefigure}

\section{Discussion}

The core formation mechanism proposed here
could begin to act even before the most massive stars
had evolved into black holes.
Evolution times for $20\msun$ stars are $\sim 8$ Myr
\citep{Schaller:92}.
The earliest phases of core formation, $\tau\lap 10^7$ yr,
would therefore be driven by the accumulation of massive stars
rather than by their remnants.
Figure 1 shows possible evidence of core growth on
time scales $\lap 20$ Myr in a few clusters.
In this context it is interesting to mention the
so-called ``young dense star clusters.''
These clusters have ages $\aplt 10$\,Myr, sizes $\sim 1$\,pc,
and contain $\aplt 10^5$ stars.
Well-known examples are NGC2070 \citep{Brandl:96},
NGC 3603 \citep{Vrba:00},
and Westerlund 1 \citep{Brandl:99}.
All of these young clusters have small but distinct cores.
The young cluster R136 in 30 Doradus ($\tau\approx 5$ Myr)
shows clear evidence of mass segregation among the
brightest stars \citep{Brandl:96}.

The mechanism described here is similar to core formation
by a binary supermassive black hole in a galactic nucleus
via the gravitational slingshot \citep{Quinlan:96}.
The latter process produces cores with masses $\sim$ a few
times the binary mass, assuming that the binary separation 
decays all the way to the point that coalescence by gravitational
wave emission can ensue \citep{Merritt:03}.
If the decay stalls at a larger separation, the displaced
mass will be smaller.
It is currently uncertain how often the decay would stall 
\citep{Milos:03}.
Core formation by a {\it population} of massive remnants
also displaces a mass that is a few
times the total mass in black holes (Figure 2), 
and because the smaller black holes are freer to move about,
there is less prospect of stalling due to a local
depletion of stars.
In galactic nuclei, the imprints left on the stellar
distribution by the clustering of stellar-mass black holes
were probably long ago erased by the growth of the supermassive black 
hole, by the formation and decay of binary supermassive
black holes during galaxy mergers, and by star formation.

A speculative application of these results is to cores
formed at the centers of dark matter halos by the clustering
of Population III remnants in the early universe
\citep{Volonteri:03}.
The latter are believed to contain at least one-half
the mass of their stellar progenitors when $m\gap 250\msun$
\citep{Fryer:01},
and the cosmological density of remnants may be similar 
to that of the supermassive black holes presently
observed at the centers of galaxies \citep{Madau:01}.
It follows that the Population III remnants could 
create cores of appreciable size, {\it if} a number of
them can accumulate in a single halo at a given time, 
and if the time for their orbits to decay is shorter 
than the time between halo mergers.
Both propositions will require further investigation.

\bigskip

DM and MH were supported by NSF grant AST02-0631,
NASA grant NAG5-9046, and grant HST-AR-09519.01-A from STScI.
SPZ was supported by the Royal Netherlands Academy of
Sciences (KNAW), the Dutch Organization of Science
(NWO), and by the Netherlands Research School for Astronomy
(NOVA).

\end{document}